 \newlength\smallfigwidth
\documentclass[aps,prb,twocolumn,floatfix,showpacs,amsmath,
amssymb,nofootinbib ] {revtex4}

\usepackage{amsmath}    
\usepackage{graphicx}   
\usepackage{verbatim}   
\usepackage{color}      
\usepackage{subfigure}  
\usepackage{float}
\usepackage{hyperref}   

\begin{document}


\title{Memristive effects in nanopatterned permalloy Kagom\'{e} array }
\author{W. B. J. Fonseca}
\affiliation{Centro Brasileiro de Pesquisas Físicas, Rio de Janeiro, Brazil}
\author{F. Garcia}
\affiliation{Centro Brasileiro de Pesquisas Físicas, Rio de Janeiro, Brazil}
\author{F. Caravelli}
\affiliation{Theoretical Division (T4), Los Alamos National Laboratory, Los Alamos, New Mexico 87545, USA}
\author{C. I. L. de Araujo}
\affiliation{Departamento de F\`{i}sica, Laborat\'{o}rio de Spintr\^{o}nica e Nanomagnetismo, Universidade Federal de Vi\c{c}osa, Vi\c{c}osa,36570-900, Minas Gerais, Brazil}

\begin{abstract}
We study memristive effects in Kagom\'{e} nanopatterned permalloy. We observe that at low frequencies a thermistor effect is present, a phenomenon arising due to the lithography and absent in similar experiments for thin films. However, we also show via an independent anisotropic magnetoresistive study that a small hysteresis accounting for 1\% of the effect is not attributable to a thermistive effect. Such effect is also confirmed by a careful subtraction scheme between nearby thermal hysteresis. In the millihertz regime, an effective model is provided to describe the experimental results for the thermistor, showing that there should be a crossover from the millihertz to the gigahertz, from a thermistor to an memresistive effect for nanopatterned permalloy.
\end{abstract}


\maketitle

\section{Introduction.}
In the last decade we witnessed an increasing interest in artificial physical systems, obtained via nanopatterning magnetic materials, as for instance artificial spin ice  ~\cite{Castelnovo1,Mengotti,topor,Chern2,Gliga}.  
 It is now possible, via novel techniques in lithographic printing, to think of tailor-designed magnetic materials addressing old and novel technological applications.

The purpose of this paper is to provide theoretical and experimental evidence for the particular application to magnetic-induced resistive memory. Memory effects in resistive materials can be used for a variety of computing applications including logical gates \cite{spin1,spin2,spin4,spin5}, unconventional computing \cite{memr1,memr2,memr3,traversa,memrc1,memrc2}, or machine learning \cite{ProbComp5,review1,review2}.

A possible pathway towards the implementation of current-controllable dynamical memory in an artificial magnetic material is via the use of anisotropic magneto-resistance (AMR) in symbiosis with many-body magnetic effects; this possibility is backed up both by recent experimental studies ~\cite{amrsi,amr,chir} and theoretical ones \cite{gwc,cargwcnis}. Instead of using disconnected magnetic nanoislands, one can focus instead on connected magnetic nanowires, which we assume to have a resistance per unit of length $\rho_0$. 

In a recent paper \cite{cargwcnis} it has been noted that the interplay between the magnetization of the permalloy and the AMR can lead to a memory effect reminiscent of the one of a memristor. Such study was also backed up recently in a tailored analysis of nanorings \cite{caravellimem}. It is thus worth exploring other mechanisms leading to similar effects in patterned magnetic devices. However, realistic devices have all sort of effects to consider, and this paper focuses on permalloy heterostructures.

A memristor is a device which satisfies Ohm's law $V=R\ I$, but moreover has a dynamical resistance of the form $\frac{dR}{dt}=f(R,I)$ (or alternatively defined in terms of the voltage). Because Ohm's law is still valid, any memory effect due to the current history in the device must still be such that to zero applied voltage there is no current leak from the device. Such requirement excludes capacitance or inductance at the first order of approximation. For instance, the Strukov-Williams memristor, initially discovered while studying Titanium Dioxide \cite{stru8}, is very well approximated by the functional form
\begin{eqnarray}
\frac{dR}{dt}=\beta I-\gamma R+\eta,R_{on}\leq  R(t)\leq R_{off}.
\end{eqnarray}
The equation above is the one of a switch, e.g. a device which changes its resistive state between two values $R_{on}\ll R_{off}$.
Such functional form has been fitted theoretically to the memristors introduced in \cite{cargwcnis}, under the assumption of spin-like islands of an artificial spin ice. The mechanism which induced the memristor feedback in the ASI approximation was a current-induced domain wall formation which is well documented in the literature \cite{spininversion1,spininversion2,spininversion3,spininversion4}. However, real magnetic materials have complicated dynamical behavior of the magnetization, many-body effects not captured by a simple Ising-like variable, and an interaction with the current captured by the Zhang-Li coupling \cite{zhangli} and temperature effects which can be attributed to thermistors \cite{thermistor}.

In the present paper, we provide theoretical, numerical, and experimental evidence that both temperature-induced and possibly magnetic memory is present in nanopatterned magnetic devices. 
In the first part of this work we discuss the experimental results.

\section{Experiments and methods}
\subsection{Sample preparation} 
\begin{figure}[!hbt]
    \centering
    \includegraphics[scale=0.4]{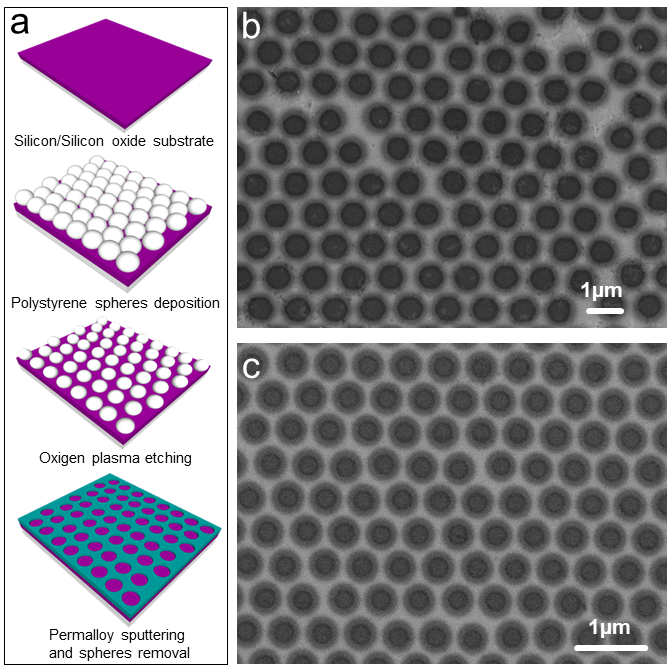}
    \caption{Colloidal lithography of the sample. We show in a) silicon substrate with 300nm thermal oxide layer, polysthyrene spheres distribution after deposition by drop casting and solution dry, spheres diameter decrease by corrosion with oxygen plasma etching and 20nm Py/3nm Pt sputtering and liftoff. Scanning electron microscopy shows the Kagom\'{e} lattices obtained by using spheres of diameter b) $1\mu m$ and better quality when spheres of diameter c) $500nm$ were used.}
    \label{fig:fig1}
\end{figure}

For the experimental verification of the memristance in nanostructured samples, we have developed Kagomé lattices by colloidal nanolithography. The process (summarized in Figure \ref{fig:fig1}a) consists of transferring floating mono-layers of self-assembled (hexagonal closely packed) polystyrene spheres (PS), at air/water interface, to a $300nm$ thermally oxidized silicon substrate \cite{Burmeister,Yi}. Then, to increase the spacing between the PS spheres, the monolayers were developed by reactive ion etching with oxygen plasma for 12.5 minutes. 

After this first step, a magnetic material deposition process was performed with $20nm$ Permalloy thin film capped by $3nm$ of platinum, deposited by sputtering over the set of substrate, plus PS spheres mask. Finally, the PS spheres were removed by deionized water ultrasonic bath to yield Kagomé lattices. Figure \ref{fig:fig1}b and \ref{fig:fig1}c shows scanning electron microscopy images of a representative lattice among the grew Kagomé lattices, using spheres with $1\mu m$ and $500nm$ diameter respectively.

 \subsection{Hysteretic Magnetoresistance}
\begin{figure}
    \centering
    \includegraphics[scale=0.28]{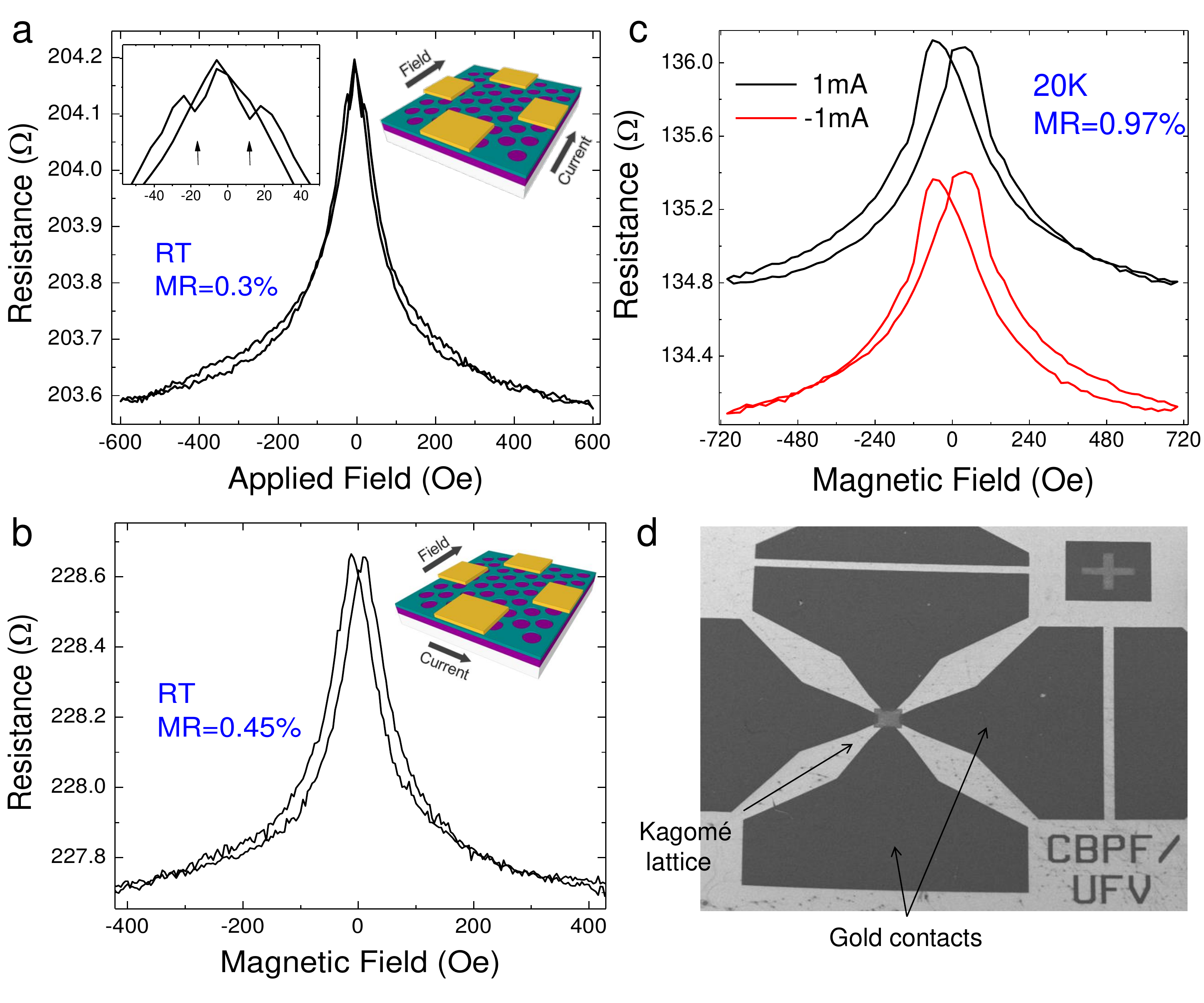}
    \caption{AMR experiment to show hysteresis in the sample. a) Magnetoresistance measured at room temperature in longitudinal configuration represented in the cartoon. The anisotropic behavior highlighted in the inset is suppressed by the effect of magnetization vertex, owing magnetoresistance signal of 0.3\%. b) Magnetoresistance measured at room temperature in transverse configuration represented in the cartoon with magnetoresistance signal of 0.45\%, here the anisotropic signal is summed with the vertex magnetization effect. c) Longitudinal magnetoresistance performed at T=20K, showing broadening of magnetoresistive peak and better observation of hysteresis asymmetry in the curve shape in function of applied current direction and d) Image of the electrical contacts utilized. }
    \label{fig:figure2}
\end{figure}

 In order to perform electric and magnetoresistive characterization of our samples, electric contacts of gold 50nm, preceded by 3nm of chrome for better adhesion, were developed by sputtering in the geometries depicted in Figure \ref{fig:figure2}.
 
 To investigate the magnetoresistive behavior we have performed measurements in the longitudinal configuration (\ref{fig:figure2}a), with an angle of $\theta$=0 between the applied current and sweep external magnetic field direction, and in the transverse configuration (\ref{fig:figure2}b) with $\theta$=90. Our results on the magnetoresistive effect are in accordance with those present in literature \cite{amrsi}. The anisotropic effect expected due to spin-orbit coupling in the Permalloy thin-film, highlighted in Figure \ref{fig:figure2}a inset, is extremely suppressed by the dynamics of the vertex magnetization, which are responsible for the isotropic behavior observed in Figures \ref{fig:figure2}a and \ref{fig:figure2}b, performed at room temperature.
 
 It is important to point out that the colloidal nanolithography we utilized has allowed an increase in the vertex size, in comparison with the nanowires width. As a result, the magnetoresistive effect we measured is approximately two orders of magnitude bigger than the ones observed in electron beam nanolithography samples of previous experiments \cite{amrsi}. Moreover, we observed in the magnetoresistive curves an even more pronounced signal in the measurements performed at $T=20$K, as presented in Figure \ref{fig:figure2}c. It was also observed an hysteresis asymmetries in the curve shape for positive and negative external fields, depending on the current signal, and in the resistance for positive and negative applied currents. This suggests that there is both the AMR effect and a pinched hysteresis in the resistivity, providing evidence a memory effect as previously predicted \cite{amrsi,cargwcnis}.
 
 
\subsection{Memristance}
\begin{figure}[hbt!]
    \centering
    \includegraphics[scale=0.28]{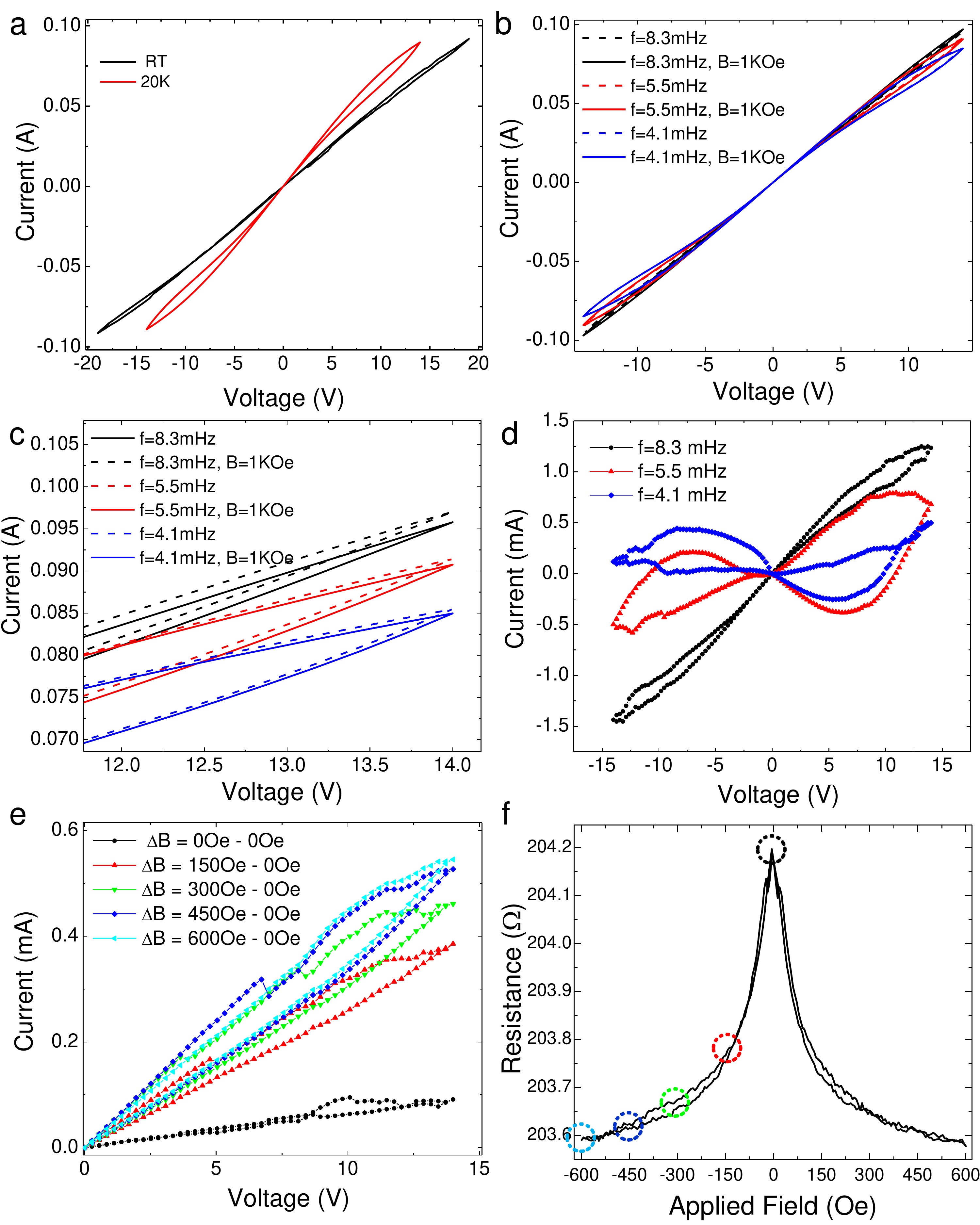}
    \caption{a) $I\times V$ curves performed at room temperature and at T=20K for Kagomé lattice and at room temperature for Permalloy thin film, b) $I\times V$ curves at T=20K with different frequency and under external magnetic field of 1000 Oe, c) zoom view of difference for the curves with and without magnetic field and d) curves obtained from the difference between measurements performed with and without external field,e) difference between measurements performed at intermediate fields and f) MR curve with marks in the fields utilized for the memristive curves.}
    \label{fig:hysteresis}
\end{figure}

To test whether the resistive memory is memristive in nature, we have performed a typical memristor experiment, in which we have checked if responding to a sinusoidal voltage input one observes a pinched hysteresis. The results obtained are presented in Fig. \ref{fig:hysteresis}. Specifically, we have performed $I\times V$ curves to investigate the resistive memory effect in our samples with currents up to $100mA$, applied in samples at room temperature and at $T=20K$. 

The results presented in Figure \ref{fig:hysteresis}a, both for a continuous Permalloy thin film and Kagomé lattice of $1 \mu m$ diameter spheres performed at room temperature and T=20K, clearly show the memristive phenomenon characterized by the typical pinched loop in the $I\times V$ curve plot. Such effect increases as temperature decreases, following the same behavior observed in the magnetoresistive curves. In the continuous Permalloy thin film, one can see that resistance is very low in comparison with Kagomé sample and just a very short loop in $I\times V$ measurement is observed, in comparison with nanowired samples (Kagomé) (\ref{fig:hysteresis}a inset).

In order to check whether the memristive hysteresis is a function of the frequency and of the external applied magnetic field, we have performed $I\times V$ curves at different frequencies under and without external magnetic field. As we show in Fig. \ref{fig:hysteresis}b, the memory effect is higher for lower frequencies, while the slight difference as a function of the external field (\ref{fig:hysteresis}c) increases as the frequency decreases. This is typical of memristive behavior.

In all the $I\times V$ measurements performed with and without an external field, the predicted resistive hysteresis was present. We noticed however memristive signals up to  13\%, much larger than the expected from the analysis of \cite{amrsi}, which considers just effects coming from many body magnetization flips, that should be in the same magnitude of sample AMR ~1\%. 

In Figure \ref{fig:hysteresis}d we plot instead the difference between the hysteresis curves, when the samples are submitted to a field strong enough to saturate the nanowires and vertices magnetization, for the different frequencies measured. Such plot is carried out in order to remove the part of signal probably coming from thermal effects, which also contribute to possible contributions coming from magnetization dynamics, leaving just the many body contribution. After the subtraction of the two curves obtained with and without magnetization saturation by external field, the characteristic pinched hysteresis loop is observed, now with the same magnitude of ~1\%. In Figure \ref{fig:hysteresis}e we present similar curves for the measurements performed under intermediate fields. A very small cycle-to-cycle variation is observed (black curve), while under successive applied field memristive effect is increased up to saturation around $600Oe$. The proportionality of the memristive signal with the resistive change under external field can be noticed in the magnetoresistive curve presented in \ref{fig:hysteresis}f.

\subsection{Thermocouple analysis}
Because of the disparity between the AMR measurement and the observed resistance change, the bulk of the memristive effect has to be found elsewhere.
It was already demonstrated in literature that persistent high currents needed to move the domain walls in such Permalloy nanowires, in same order of $1$ x $10^{12}A/m^2$ utilized in the present work, could be able to increase the temperature up to $800K$ \cite{yamaguchi} in nanowires with similar geometries to those utilized in our samples. Such temperatures could be high enough to stimulate thermal fluctuations of magnetization \cite{renan}.
 \begin{figure}[hbt!]
     \centering
     \includegraphics[scale=0.27]{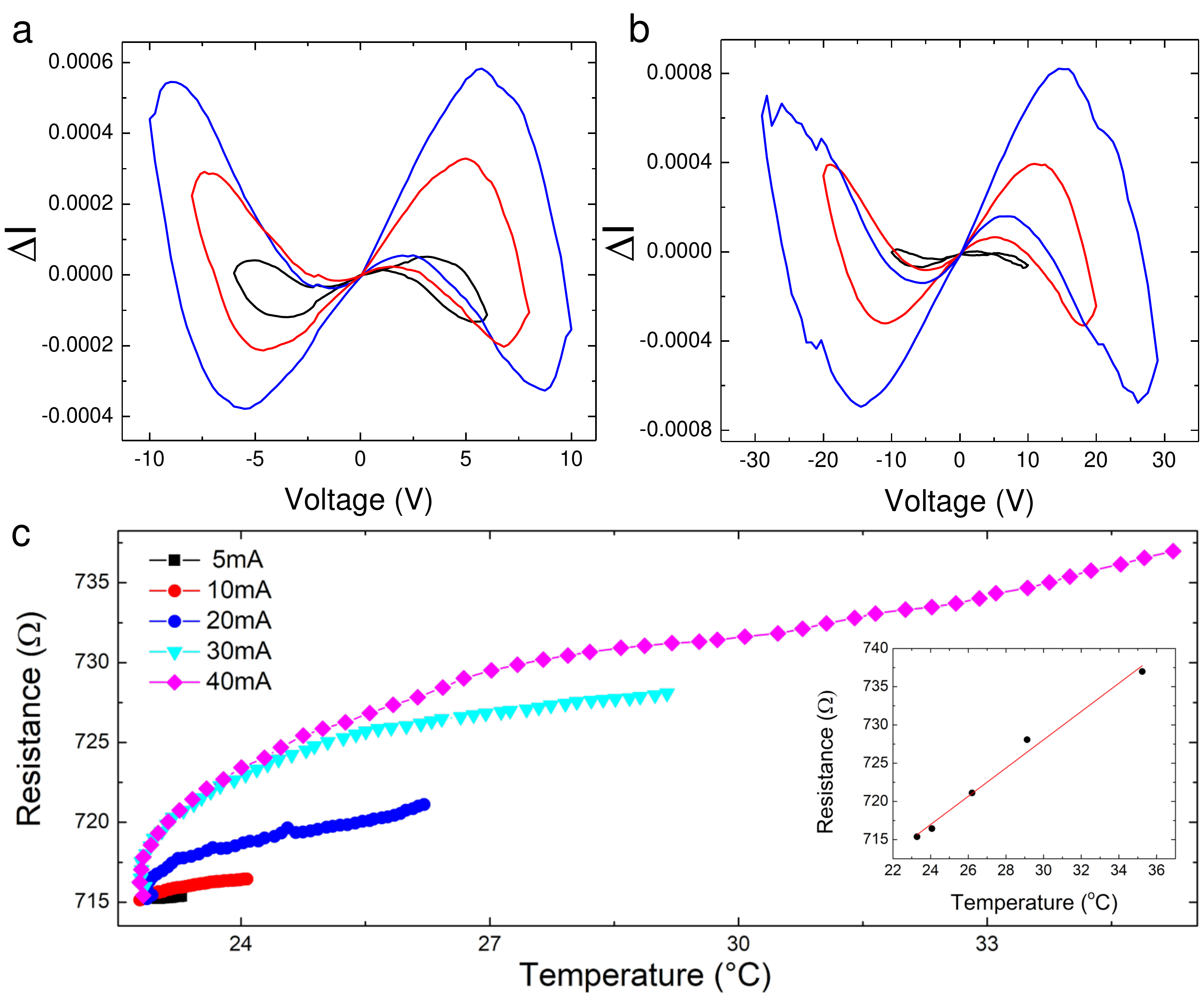}
     \caption{Residual linear fit of $I \times V$ curve performed with different applied voltages in  Kagomé lattice from spheres of a) 0.5 $\mu$m and b) 1 $\mu$m. c) Sample resistance and temperature variation in function of different currents applied for 50s interval.}
     \label{fig:thermal}
 \end{figure}
 
An insight of thermal effects relevance in the high memristive signal measured can be obtained from the measurements performed in samples with different nanowire lengths presented in Figure \ref{fig:thermal}. The resistance measured in a sample obtained with $500 nm$ diameter spheres (\ref{fig:thermal} a) is half of sample resistance obtained with $1 \mu m$ spheres (\ref{fig:thermal} b), which presents higher memristive effects with same applied current of $100 mA$.

To map the evolution of resistance in function of sample heating by Joule effect, we have applied different persistent currents and measured the evolution of sample temperature with a K-type thermocouple for $50s$, timescale similar to the utilized for the $I\times V$ curves. Results presented in (\ref{fig:thermal} c) show linear evolution of resistance of around $0.54 \Omega$ per Celsius degree.      

The difference between the curves of Fig. \ref{fig:thermal} (c), and the hysteretic AMR analysis previously  discussed, show that a temperature-based memristor model alone cannot explain all the effects in the material.

\section{Theoretical analysis of temperature contributions}
\subsection{Thermistor effect in heterostructures}

While the splitting of the AMR experiment confirms a spin transfer torque effect, at the Hertz frequency we still have an effect of the temperature on the resistivity, causing a memristive hysteresis \cite{thermistor}. It is worth here to characterize the properties of this thermistor in our samples. Let us assume that the resistivity of the material changes with the temperature, and that moreover  we are in the regime of a thermistor. We thus ignore for the moment the AMR effect, e.g. a temperature dependent resistor with conductance $R(T)$. We want to show here that such effect is important because of the nanopatterning, which is consistent with the experimental results, due to the difference in area between a thin film and an heterostructure.

In order to explain this effect we used a thermistor model \cite{thermistor} for the material. This is given by
\begin{eqnarray}
V(t)&=&R(w_T) I(t)\\
R(w_T)&=&R_0 \Big(1+T_r \alpha_T (w-w_0)\Big)\\
\frac{dw_T}{dt}&=&\frac{V(t)^2}{\tau_v \bar v^2 (1+T_r \alpha_T( w_T-w_0))}-\frac{1}{\tau}(w_T-1)
\end{eqnarray}
where $T_r$ is room temperature, and for permalloy we have $\alpha T_r\approx \frac{1}{3}$.
The parameter $w_T$ physically represents the internal temperature of the permalloy. The parameter $R_0$ is fit to be approximately $720 \Omega$ for our device at room temperature. The parameter $\tau\approx 1-10$ seconds and represents the thermalization time of the sample at room temperature.

Then, Ohm's law for the effective resistance of the sample follows 
 \begin{eqnarray}
 V(t)= R_{eff}(t,T) i(t).
 \end{eqnarray}
 The sample can dissipate energy both through the silicon substrate and via radiative transfer.
 If $\delta$ is the conduction per unit of area, $\sigma$ the Stefan-Boltzmann constant, and $c$ the heat capacity of the material, we have
 \begin{eqnarray}
 C\frac{dT}{dt}&=&i(t) v(t)-A\Big( \delta (T-T_0)+\sigma (T^4-T_0^4)\Big) \nonumber \\
 &=&i(t)^2 R_{eff}(t,T)-A\Big( \delta (T-T_0)+\sigma (T^4-T_0^4)\Big)\nonumber \\
 \end{eqnarray}
 and thus we obtain an effective memristive effect. 

  In order to estimate the frequency at which the thermistor effect is negligible, we use the the equation above. The estimate is independent from many body effects, and since the effective resistance of the sample depends on the resistivity linearly (while not on the topology) we can work with an effective model. The specific heat capacity of permalloy is $c_v=0.124$ cal/g/Celsius, the density $\rho=8.74 g/cm^3$. Also, it has been known for some time \cite{soviets} that the electrical resistivity of permalloy is linear in temperature in the range $273-1000$ Celsius, and nearly doubles in that interval (80\% gain in resistivity).
  However, we are not aware of resistivity measurements for structures like ours.  
  Given a certain initial condition, $R_0$, thermocouple experiments are consistent with a negative coefficient thermistor, which follows an equation
  \begin{eqnarray}
  R(T)=R(T_0) e^{-C(\frac{1}{T}-\frac{1}{T_0})}.
  \end{eqnarray}
  with $C$  
  The total heat capacity can be written as $C=c_v\times V \times \rho$, where $\rho$ is the density of the sample, and $V=A\times h$ is the volume, given by the area times the thickness. 
  
  In our experiment we can have estimated that $R_0\approx 720 \Omega$ at $T_0=307 K$, and $R_1\approx 755 \Omega$ at $T_1=355 K$ , from which we can estimate $\alpha \approx 10^{-3} K^{-1}$. Also, we neglect the radiative transport, since it is orders of magnitude smaller than the dissipation due to transfer of energy via the plate.

 In  order to make the equation adimensional, we divide by the room temperature energy, $E_r=\kappa T_{r}$. If we define 
 \begin{eqnarray}
 w_T=\frac{T}{T_{r}},
 \end{eqnarray}
 we can write 
  \begin{eqnarray}
 \frac{\partial w_T}{\partial t}=\frac{  i(t) v(t)}{ C T_r}-\frac{A\delta }{  C} (w_T-w_0) \nonumber \\
 \end{eqnarray}
and writing explicitly $C$ and $i(t)$ in terms of the area and the resistance explicitly, we obtain
  \begin{eqnarray}
 \frac{\partial w_T}{\partial t}&=&\frac{   v^2(t)}{A c_v m h T_r R_0 (1+T_r \alpha_T(w_T-w_0)) } \nonumber \\
 &&-\frac{\delta }{  c_v h \rho} (w_T-w_0) ,
 \end{eqnarray}
and where we used
\begin{eqnarray}
R(w_T)=R_0(1+ T_r \alpha_T (w_T-w_0)),
\end{eqnarray}
with $w_0=\frac{T_0}{T_r}$; thus, if we choose $T_0$ to be room temperature, then $w_0=1$. For practical purposes, we can set $T_r \alpha\approx 0.3$. Now, we note that $\frac{1}{\tau}=\frac{\delta }{c_v h \rho}$ is a relaxation timescale, while
$A c_v \rho h T_r R_0=\tau_v^{-1} \bar v^2$ is an effective activation voltage per unit of time, which depends on the area of the permalloy sample in contact with the surface. This implies that the smaller the area the more pronounced the memristive effect due to the temperature will be. This is one of the reasons why this effect is typically negligible in thin films, while it is more pronounced in our sample.
Let us assume that the thermistor is near the equilibrium. What we are interested in is the interplay between the decay and the forcing and the relationship in terms of $\omega$. We thus neglect the $w$ dependence on the denominator of the voltage-forcing term.

The solution of the differential equation in this regime, and assuming that $v(t)=V_0 \cos(\omega t)$ is given by
\begin{eqnarray}
w_T(t)&=&c_1 e^{-\frac{t}{\tau }} \nonumber \\
&+&\frac{2 \tau ^2 V_0^2 \omega  \sin (2 t \omega )+\tau  V_0^2
   \cos (2 t \omega )}{2 {\bar v}^2 \left(4 \tau ^2 \omega ^2+1\right)}\nonumber \\
&+&\frac{\tau  V_0^2
   \cos (2 t \omega )+\left(4 \tau ^2 \omega ^2+1\right) \left(\tau  V_0^2+2 {\bar v}^2
   w_0\right)}{2 {\bar v}^2 \left(4 \tau ^2 \omega ^2+1\right)}\nonumber \\
\end{eqnarray}
where $c_1$ is constant of integration given the initial condition, which however we see goes to zero exponentially fast. 
As a result, from the oscillations around equilibrium, we can estimate in a linear response framework the value of $\bar v$.
 Since in our experiments we do not see a strong decay or thermalization, we expect $\tau$ to be of order $\tau \sim 1-10$ seconds. Let us thus assume first that $\tau\omega \ll 1$. In this case, we have
\begin{eqnarray}
\Delta w_T(t)=\frac{ V_0^2}{\tau_v \bar v^2}  \omega \tau^2 \sin(2 \omega t).
\end{eqnarray}
Since we observe that the oscillations in resistance are of order $\Delta R/R\approx 0.03$, we obtain a rough estimate for $\bar v$ in the mhz of
$\frac{\Delta w}{3}\approx 0.03$ at $V_0=10 V$. We thus have
\begin{eqnarray}
\frac{V_0^2 \omega \tau^2}{3\times 0.03}\approx \tau_v \bar v^2.
\end{eqnarray}
Substituting these, we have 
\begin{eqnarray}
\tau_v \bar v^2\approx 11.1\ [V]^2[s],
\end{eqnarray}
which, setting $\tau_v=1$ as a unit of time, leads to an effective voltage necessarily to observe a noticeable response in the frequency regime of our experiments of $\bar v= V_0\approx 3.33$ Volts. This is indeed consistent with the experimental results.

In the regime, $\omega\gg \tau^{-1}$, 
we have, up to the first order in $1/\omega$
\begin{eqnarray}
w_T(t)=\frac{\tau V_0^2+ 2 \bar v^2\tau_v w_0}{2 \bar v^2 \tau_v}+\frac{1}{\omega} \frac{2 \tau^2 V_0^2 }{2 \bar v^2 \tau_v \tau^2}\sin(2 t\omega)+O(\frac{1}{\omega^2})\nonumber 
\end{eqnarray}
from which we see that while the equilibrium depends on $\tau$, the oscillations do not.  We obtain
\begin{eqnarray}
\Delta w_T(t)=\frac{1}{\omega} \frac{ V_0^2 }{ \tau_v \bar v^2 }\sin(2 t\omega).
\end{eqnarray}
The result above means that if we increase the frequency $\omega$ by $K$, these should be suppressed by a factor $\frac{1}{K}$. 
This argument suggest that if we observe a memristive effect in the $centihz$ or $millihz$ due the thermal changes, these should be suppressed by a factor $10^{-8}:10^{-9}$ in the Ghz, and thus negligible.

As a last comment, we now discuss why in thin films this effect is not usually observed, as we also found out in our experiments. The surface of the sample is roughly 25 $mm^2$, or $25\times 10^{-6} m^2$. We have prepared two samples, one with sphere of $1\mu m$ and $500 nm$ diameter. In the $500 nm$ sample, we can perform the calculation by directly assuming a certain density of holes. In an area of  $A_f=2.3*10^{-11}$ $m^2$ we have roughly 110 spheres, give or take. This implies that we have itched out from the surface an area $A_{i}=110*\pi*(0.25*10^{-6})^2$, and thus we obtain a remaining area  $A_s=A_f-A_i=1.4*10^{-12}\ m^2$. The ratio between $A_f/A_s\approx 16$. This implies that the ratio between the effective activation voltage per unit of second of the thin film is given by
\begin{equation}
    \frac{{\bar v}^2_{itched} }{{\bar v}^{2}_{thin\ film}}=\frac{1}{16},
\end{equation}
or $4\ \bar v_{itched}= \bar v_{thin\ film}$. The effect is thus at least one of magnitude smaller in a thin film permalloy, which implies voltages of at least $\approx 30$ Volts to fully appreciate the memory effect, which is prohibitive without burning the sample.

In conclusion of the model above, we obtained that the thermistor controlled at a certain frequency $\omega$ has two regimes for the change of resistance per unit of time (second):
\begin{itemize}

    \item $\omega \tau \ll 1$: the time evolution of the system leads to a oscillatory regime around the mean, with $$\Delta R/T\approx \frac{R_0 V_0^2}{\tau_v \bar v^2} \omega \tau^2 \sin(2\omega t)$$
    \item $\omega \tau\gg 1$: a solution of the dynamics leads to a change in amplitude, given by $$\Delta R/T\approx \frac{R_0 V_0^2}{\tau_v \bar v^2  \omega}  \sin(2\omega t)$$
\end{itemize}
Again, since in our experiments we do not observe a strong decay or relaxation, we expect the relaxation time to be of the order of the seconds. We are thus in the first of the two regimes, which is what we use to estimate $\tau_v \bar v^2$, which is approximately of $3.3$.  Thus, we obtain the result that at slow frequencies the system is dominated by thermal effects. This is one of the reasons which explains why in our experiments we do have little dependence of the hysteresis curve on an applied field. Yet, as we show above, for larger frequencies resistance memory oscillations are suppressed by a factor $\frac{1}{\omega}$. This implies that at higher frequencies we have a crossover effect between the thermal and the magnetic degrees of freedom in the material, which we characterize next.

\subsection{Magnetization dynamics} 
In a previous work it was argued that when a current is run through the Kagomé permalloy, the total resistance is finely controlled by the direction of magnetization of the wires~\cite{amrsi,amr,gwc}.
It was shown \cite{gwc} that the effect of the magnetoresistance is due to the presence of the vertices, where domain walls form, and the system can thus be interpreted as an electrical circuit with voltage drops at the vertices. In \cite{cargwcnis} it has been noted instead that such construction can be mapped to a resistor network with voltage generators in series, and thus can be written in an analytical form  and mapped to an effective memristive effect.

More recently, it has been shown that in magnetic nanorings, the interplay between AMR and the Zhang-Li coupling are sufficient for the ring to exhibit memristive effects which are strongly dependent on the internal magnetization states \cite{caravellimem}.
As we have seen here, magnetic permalloy patterned as Kagomé lattice shows a residual memristive effect (on top of the bulk thermistor effect) which is magnetic in nature.
We wish to show here that such effect can in principle be attributed to a many body phenomenon which we attempt to qualitatively account for here. 

As discussed in \cite{cargwcnis},
for heterostructures such as a Kagomé patterned permalloy, the interplay between the AMR and the local magnetization of the structure can affect the resistivity of the material. Here we go beyond the spin-like approximation used in \cite{cargwcnis} and study the magnetization reversal in a realistic model.
Specifically, magnetization reversal processes in Kagom\'{e} nanowires were verified by micromagnetic simulations based on the open-source GPU-based software $MUMAX^3$.

From energy minimization, one obtains sequences of snapshots between spin configurations, which are based on the Landau-Lifshitz-Gilbert (LLG) equation with the spin-transfer torque,
\begin{eqnarray}
\frac{\partial\boldsymbol{M}}{\partial t}=\gamma\boldsymbol{H}_\text{eff}\times\boldsymbol{M}
+\frac{\alpha}{M_\text{S}}\boldsymbol{M}\times\frac{\partial\boldsymbol{M}}{\partial t}
-u\frac{\partial\boldsymbol{M}}{\partial y}\nonumber\\+\frac{\beta}{M_\text{S}}\boldsymbol{M}\times\frac{
\partial\boldsymbol{M}}{\partial y},
\label{eq:LLG}
\end{eqnarray}
In simulations we have used periodic boundary conditions on a lattice of $5\mu$m $\times$ $5\mu$m size and with finite difference discretization for the iterations (following Eq.~(\ref{eq:LLG}) with cubic cell of $5$ nm $\times$ $5$ nm $\times$ $5$ nm). The numerical parameters we used were those of permalloy, for which later in this letter we performed the experiments. They were magnetic saturation $M_\text{S}=860\times 10^3$ A m$^{-1}$, exchange constant $A_\text{ex}=13\times 10^{-12}$ J m$^{-1}$, polarization $P=0.5$ and Gilbert damping $\alpha=0.01$ for dynamics. In simulations we introduced a density of current $J_\text{y}=2\times 10^{13}$ A/ m$^{2}$, cycling at a frequency $\omega$. The numerical calculations were carried by a second-order Heun’s solver with a fixed time step of $1\times10^{15}s$ \cite{Leliaert}.

\begin{figure}[hbt!]
    \centering
    \includegraphics[scale=0.25]{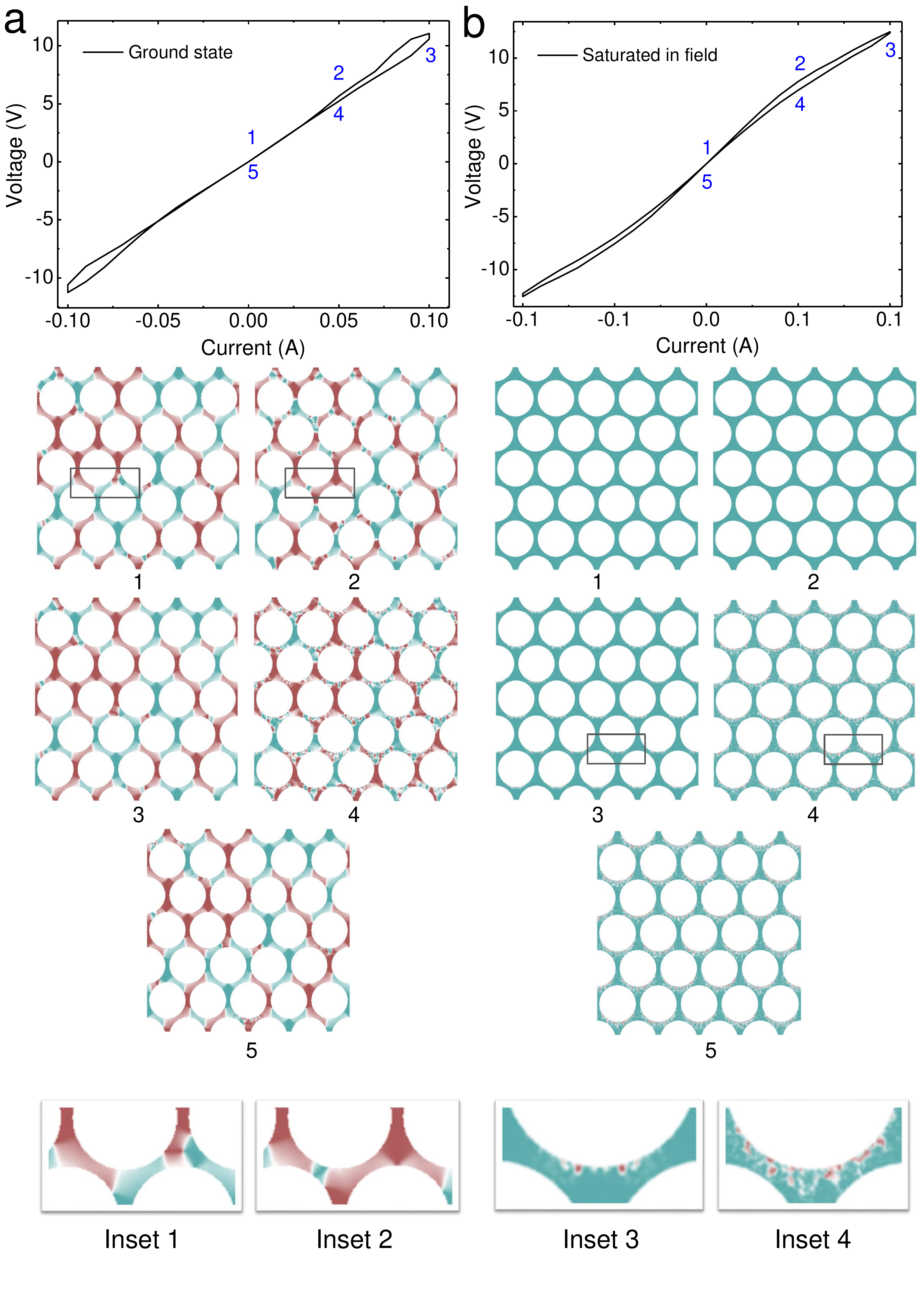}
    \caption{a) $I\times V$ curves obtained from micromagnetic simulation performed by spin polarized current sweep in a sample with ground state magnetization. Frames of magnetization states in points 1-5 of the loop are presented with zoom view in insets 1 and 2. b) Same procedure performed in a sample with magnetization saturated by a magnetic field of 0.3T.}
    \label{fig:figmicro}
\end{figure}

To better understand the role of magnetization dynamics in the memristive effect in the $I \times V$ measurements performed in kagomé lattices, while an  analysis of extended magnetic objects escapes a complete analytical treatment, the intuition for the behavior of electrical-magnetic interactions can be obtained via intense numerical scrutiny. For that purpose we have utilized micromagnetic simulations in samples without (\ref{fig:figmicro}a) and with (\ref{fig:figmicro}b) the presence of an applied 0.3T constant magnetic field on the y-axys.

In the frames of magnetization distribution taken in positions 1-5 during the current loop, one can see that in the lattice without external magnetic field two processes are present, domains wall motion by the current, which is applied in the y-direction, with nanowires magnetization flip in the x-direction (inset 1 and 2), aside of periodic domain wall creation and annihilation by the particular geometry of the borders for high currents\cite{sitte}. 

Despite the many-body effect contribution to the memristive effect, analytically predicted in previous work \cite{amrsi}, the asymmetric disorder brought by the periodic domain wall creation between increasing (frame 2) and decreasing (frame 4) current regime is mainly responsible for the memristive effect observed. In the saturated regime under an external magnetic field, the many-body effect is not present as expected, however, the geometric anisotropy is strong enough to shift the magnetization in relation to the current polarization direction to achieve the condition for periodic domain wall creation, then even in saturation regime, the asymmetric disorder is observed (insets 3 and 4).            
 
\section{Conclusion}
The present paper is the first experimental investigation of the AMR-induced memristive effect in magnetic materials.

We have lithographically printed (via colloidal deposition) a Kagom\'{e} lattice of permalloy nanowires with two different lengths, and studied the device memristive phenomenology in comparison with the thin film. The history dependence of the magnetization on the current has been observed in the analysis of the anisotropic magnetoresistance, which has a higher signal in comparison with e-beam lithographed samples, due to the large area of vertex obtained. While the bulk of the phenomenon can be explained via a thermistive effect not present in thin films, also confirmed via a thermocouple analysis, we have found that 1\% of the effect can be associated to the AMR induced hysteresis. 

In order to characterize the phenomenon, we have studied via micromagnetic simulations the effect of the current in the sample, and found that domain walls and spin waves can be found. However, thermal and dynamicals effects could be removed by analyzing the difference between curves obtained in the experiments with and without external fields, and the many body memristive characteristic could be measured.

While overall we do find that a residual effect can be attributed to a magnetic-induced memristive effect,  both in numerical simulations and experiments, the presence of the contribution of a thermistive effect prevents us for making precise statements on the nature of the phenomenon. This said, we do find good evidence that the combination of a coupling between the magnetization and the current via spin-torque and anisotropic magneto-resistance leads to a polar memristive effect. Some comments are in order. It is interesting to note that the hysteresis curves of our experiments exhibit very little noise levels compared to other similar devices.\cite{ProbComp5} This is typically not the case for memristors, which have (again, typically) noise in the hysteresis curve.

Our numerical results and experiments suggest that tuning parameters, and using different materials,  could allow domain wall motion at lower currents.
The theoretical analysis we performed on the thermistor also suggest that the thermistive effect should not be present at higher frequencies. Thus, experiments performed in the Ghz regime should eliminate spurious effects. Similarly to what described for the case of magnetic rings \cite{caravellimem}, pure magnetically induced memristive effects should be present at that scale.

\section{Acknowledgements.} 

The financial support for this research was provided by the Brazilian agencies FINEP, FAPEMIG, and CAPES (Finance Code 001). The work of FC was carried out under the auspices of the US Department of Energy through the Los Alamos National Laboratory, operated by Triad National Security, LLC (Contract No. 892333218NCA000001).

\section*{Author contributions} 
W.B.J.F and F.G. designed and developed the samples. C.I.L.A performed the experimental measurements and numerical simulations and F.C worked on the theoretical model. All authors contributed to discussing the experiments, the models to explain it and writing the paper.

\end{document}